\documentclass[aps,pra,twocolumn]{revtex4-2}
\usepackage{graphicx}
 \usepackage{lmodern}
 \usepackage{amsmath,amssymb,bm,amsthm,mathtools}

\begin{document}
\newtheorem{theorem}{Theorem}
\newtheorem{corollary}{Corollary}
\newtheorem{proposition}{Proposition}
\newtheorem{lemma}[proposition]{Lemma}
 \theoremstyle{remark}
\theoremstyle{plain}
\newtheorem{definition}{Definition}
   
\def\be{\begin{equation}}
\def\en#1{\label{#1}\end{equation}}
 \def\S{\mathcal{S}}
\def\D{\mathcal{D}}
 \def\vare{\varepsilon}
\newcommand{\per}{\mathrm{per}}
\newcommand{\one}{\boldsymbol{1}}

\title{  Marginal bunching  for   \textit{in situ}  certification of  multiparticle  indistinguishability    }  
\author{Valery Shchesnovich }
\affiliation{Centro de Ci\^encias Naturais e Humanas, Universidade Federal do
ABC, Santo Andr\'e,  SP, 09210-170 Brazil }
\date{\today}

\begin{abstract}
We propose an \textit{in situ} method for certifying the multiparticle indistinguishability of identical bosons or fermions in generic linear interferometers. For particles occupying distinct input ports, the method uses two-particle marginal bunching or anti-bunching probabilities to bound full multiparticle indistinguishability through Aldous’ spectral gap theorem. For Haar-random unitary interferometers, the number of experimental runs required to estimate this bound to a prescribed precision, at fixed confidence, is polynomial in the particle number, the interferometer size, and the inverse precision, with high probability over the choice of interferometer. Losses are incorporated without conditioning on particle survival. The method is particularly suited to certifying photon indistinguishability in large-scale interference experiments, including Boson Sampling experiments targeting quantum computational advantage.
 
  \end{abstract}
\maketitle	

\section{Introduction}

Multiparticle indistinguishability is a key resource for photonic quantum information processing. It underlies the interference exploited in Boson Sampling~\cite{AA}, demonstrated with large numbers of single photons~\cite{ExpBS2019}, and in Gaussian Boson Sampling experiments~\cite{ProgGBS2022,Jiuzhang4}. It is also central to quantum computation with linear optics~\cite{LOCC2001}, high-fidelity photonic logic gates~\cite{PhotLogGates2022}, the generation of photonic cluster states~\cite{PhotClusterSt2023}, and programmable photonic computing platforms~\cite{PrototPhotComp2024}. Partial distinguishability can enable efficient approximate classical simulation of Boson Sampling~\cite{EffClassBS2018,RenShchPatr2018,Moylett2020}. Reliable assessment of indistinguishability is therefore important both for developing photonic quantum technologies and for evaluating their computational capabilities.

The Hong--Ou--Mandel effect~\cite{HOM}, in which two indistinguishable photons entering distinct input ports of a balanced beam splitter leave through the same output port, provides a basic test of two-photon indistinguishability. Similarly, fermionic anti-bunching reflects fermionic exchange statistics~\cite{2FAntiBunch}. Hong--Ou--Mandel interference has been used to characterize the indistinguishability of consecutively emitted photons~\cite{quantDot2002}. Related methods also allow single-photon spectral purity to be inferred from the temporal width of the interference dip together with the photon spectrum~\cite{cassemiro2010}. Extending such diagnostics to many particles is less straightforward. For example, recent four-photon experiments show that high interference contrast need not imply a correspondingly high degree of photon indistinguishability~\cite{4phHOM2026}.

Assessing multiparticle indistinguishability in a large interferometer is challenging~\cite{ExpVerBS2014}. Individual output probabilities are  exponentially small, thus  computing the full output distribution is generally intractable in the regime targeted by Boson Sampling~\cite{AA}. Full bunching into a single output port provides a particularly simple diagnostic~\cite{1portBunch2013}, but, for particles occupying distinct input ports, even its maximal probability decreases exponentially with particle number~\cite{Bunch1FourExp2025}. Several approaches avoid relying on individual rare events. Suppression laws in Fourier interferometers provide efficient tests of multiparticle interference~\cite{FourZT2014}. Tight bounds on multiparticle indistinguishability have also been derived from a two-particle symmetry measure accessible through a polynomial number of observables~\cite{NewDist2024}. Bunching and photon-number variance have been used for semi-device-independent characterization~\cite{Bunch1FourExp2025}, while recent Fourier-based protocols provide scalable certification under specified assumptions on the input states and measurement settings~\cite{FourImpr2026,FourIntProt2026}. Bayesian and maximum-likelihood approaches offer complementary assessments based on models of partial distinguishability~\cite{Bayes2023,ML4BS2021}. An important goal is to perform certification directly in the interferometer used for the target experiment.

Generalized boson bunching into a prescribed subset of output ports~\cite{GenBunch2016} was proposed as a tool for assessing Boson Sampling. Its use for indistinguishability certification, however, requires identifying when the bunching probability is maximized only by completely indistinguishable bosons. Counterexamples to permanent inequalities~\cite{POT2016,tran2022,Drury2016} rule out an unrestricted maximality claim. The corresponding enhancement of bunching by partially distinguishable bosons is known as anomalous bunching~\cite{AnBunch2023,ExpAnBunch2026}. Maximality of generalized bunching has been established after Haar averaging and linked to a conjecture concerning states invariant under permutations of occupied input modes~\cite{ConjBunch2026}. These results leave open the general problem of obtaining useful indistinguishability certification guarantees from generalized bunching in a fixed interferometer.

Here we introduce marginal bunching as a basis for \textit{in situ} certification of multiparticle indistinguishability in any  generic linear interferometer. The method applies to both bosons and fermions, allows arbitrary internal states, including correlations, and also addresses the emulation of one species by the other. We establish certification guarantees based on the extremal properties of two-particle marginal bunching and prove an analogous maximality result for three-boson marginals. For particles occupying distinct input ports, we use Aldous' spectral gap theorem to convert two-particle marginal bunching or anti-bunching probabilities into efficiently computable bounds on full multiparticle indistinguishability. For Haar-random unitary interferometers, the number of experimental runs required to estimate these bounds to a prescribed precision, at fixed confidence, is polynomial in the particle number, interferometer size, and inverse precision, with high probability over the choice of interferometer. Losses are incorporated directly into the detected marginals, without conditioning on particle survival. Under uniform losses, this polynomial scaling is preserved whenever the inverse single-particle survival probability is polynomially bounded.

The paper is organized as follows. Section~\ref{Sec2} establishes the two-particle bunching and anti-bunching properties underlying the method. Section~\ref{Sec3} introduces the multiparticle framework and defines $r$-particle marginal bunching for $2\le r\le n$, with full generalized bunching recovered for $r=n$. Section~\ref{Sec4} proves the extremal properties of two-particle marginal bunching and their implications for full multiparticle indistinguishability. Section~\ref{Sec5} extends the maximality result to three-boson marginals. Section~\ref{Sec6} derives the quantitative certification bounds and analyzes their sampling cost. Section~\ref{Sec7} summarizes the results and discusses directions for further research. Technical proofs and supporting derivations are provided in Appendices~\ref{appE}--\ref{appD}.

\section{Two-particle bunching into a subset of output ports}
\label{Sec2}

We begin with two identical particles, either bosons or fermions, interfering in a linear interferometer described by the matrix $U_{kj}$, $k,j=1,\ldots,M$. Losses are included by allowing $UU^\dag\le I$. We assume that  the internal degrees of freedom of the particles are invariant under the action of interferometer.

We select a nonempty proper subset $\Omega_K$ of the output ports, with $1\le K<M$. For simplicity of notation, let  $\Omega_K=\{1,\ldots,K\}$. The probability that a particle entering input port $k$ is detected in this subset is
$H_{kk}=\sum_{j=1}^K|U_{kj}|^2$.
For two particles entering distinct input ports $k\ne l$, their interference depends on the state of their internal degrees of freedom, hereafter called the label state. For example, for independently prepared particles, this state has the product form
$\hat{\varrho}^{(L)}=\hat{\rho}_1\otimes\hat{\rho}_2$,
where $\hat{\rho}_1$ and $\hat{\rho}_2$ are the individual label states. The probability of detecting both particles in $\Omega_K$ is
\begin{eqnarray}
&&P(k,l)=H_{kk}H_{ll}\pm
\mathrm{Tr}(\hat{\rho}_1\hat{\rho}_2)|H_{kl}|^2,
\label{pkl}\\
&&H_{kl}:=\sum_{j=1}^K U_{kj}U^*_{lj},\nonumber
\end{eqnarray}
where the upper sign corresponds to bosons and the lower sign to fermions. Since $0\le H\le I$, the extremal expressions satisfy
$0\le H_{kk}H_{ll}\pm|H_{kl}|^2\le1$.

For a beam splitter with nonzero transmission and reflection, selecting a single output port $j$ gives
\[
P(1,2)=
\left[1\pm\mathrm{Tr}(\hat{\rho}_1\hat{\rho}_2)\right]
|U_{1j}|^2|U_{2j}|^2.
\]
This probability is maximized for bosons and minimized for fermions precisely when
$\mathrm{Tr}(\hat{\rho}_1\hat{\rho}_2)=1$.
For independently prepared particles, this condition requires identical pure label states, hence the particles are indistinguishable.

For a general  label state $\hat{\varrho}^{(L)}$, Eq.~(\ref{pkl}) remains valid with
$\mathrm{Tr}(\hat{\rho}_1\hat{\rho}_2)$ replaced by
\[
J_{12}:=\mathrm{Tr}\!\left(\hat{P}_{12}\hat{\varrho}^{(L)}\right),
\qquad -1\le J_{12}\le1,
\]
where $\hat{P}_{12}$ exchanges the two internal tensor factors. The condition $J_{12}=1$ is equivalent to the label state being supported on the symmetric subspace and expresses two-particle indistinguishability. We therefore obtain the following result.

\begin{proposition}[Two-particle bunching/anti-bunching]
\label{prop1}
Consider two identical bosons or fermions occupying distinct input ports $k$ and $l$ of a fixed linear interferometer. For a fixed  proper subset of output ports satisfying $H_{kl}\ne0$, the probability of detecting both particles in that subset attains its maximum for bosons, or its minimum for fermions, if and only if the particles are   indistinguishable, i.e., $J_{12}=1$.
\end{proposition}

Indeed, the probability depends linearly on $J_{12}$ with a nonzero coefficient, so the relevant extremum occurs only at $J_{12}=1$.

The multiparticle method extends this criterion to an input of $n$ particles. Two-particle marginal bunching probabilities can be extracted from the full experimental output records and combine the interference contributions of different particle pairs. Under suitable genericity conditions, their extremal values enforce pairwise symmetry, which in turn implies full multiparticle indistinguishability. This provides the basis for certification \textit{in situ} in the interferometer used for the experiment.

\section{Indistinguishability, marginal probabilities and marginal bunching}
\label{Sec3}

To develop the certification method, we first recall the description of multiparticle indistinguishability and two equivalent expressions for marginal probabilities. We then introduce marginal bunching.

\subsection{General $n$-particle input state and output probability}
\label{Sec3A}

Consider $n$ identical particles entering a linear interferometer with input occupation vector
$\bm{n}=(n_1,\ldots,n_M)$,
where $|\bm{n}|:=n_1+\ldots+n_M=n$.
The probability of detecting all $n$ particles with output occupation vector
$\bm{m}=(m_1,\ldots,m_M)$, $|\bm{m}|=n$, is
\cite{IndTh2015,GenBunch2016,IndTh2026}
\be
p_{\bm{m},\bm{n}}\! = \!\frac{1}{\bm{m}!\bm{n}!}\!
\sum_{\sigma,\pi\in S_n}\vare(\sigma\pi)J(\pi\sigma^{-1})\!
\prod_{\alpha=1}^n
U_{k_{\sigma(\alpha)},j_\alpha}
U^*_{k_{\pi(\alpha)},j_\alpha},
\en{p{m,n}}
where $\bm{m}!:=m_1!\ldots m_M!$, and similarly for $\bm{n}!$.
Here $\vare(\sigma)=1$ for bosons and
$\vare(\sigma)=\mathrm{sgn}(\sigma)$ for fermions.
The ordered sequences $k_1\le\ldots\le k_n$ and
$j_1\le\ldots\le j_n$ encode the input and output occupations, respectively.

The dependence on the particles' label state
$\hat{\varrho}^{(L)}$ is encoded by the indistinguishability function on the symmetric group $S_n$,
\be
J(\sigma):=\mathrm{Tr}(\hat{P}_\sigma\hat{\varrho}^{(L)}),
\quad
\hat{P}_\sigma\bigotimes_{\alpha=1}^n|\phi_\alpha\rangle
:=
\bigotimes_{\alpha=1}^n|\phi_{\sigma^{-1}(\alpha)}\rangle.
\en{J}
Subject to the symmetry constraints imposed by the input occupations, $\hat{\varrho}^{(L)}$ may be an arbitrary density operator on the tensor-power Hilbert space of the internal degrees of freedom. In particular, it may describe an entangled state that allows particles of one species to emulate the other
\cite{walkEntPhot,BSF2015}.

For particles occupying the same input port, the label state must be supported on the symmetric subspace for bosons or the antisymmetric subspace for fermions. Consequently, for every $\sigma\in S_n$,
\be
J(\nu\sigma)=J(\sigma\nu)=\vare(\nu)J(\sigma),
\quad
\forall\nu\in S_{n_1}\times\ldots\times S_{n_M},
\en{JYoung}
as required by the bosonic or fermionic commutation relations
\cite{IndTh2015,GenBunch2016,IndTh2026}.

For particles emitted by independent sources into distinct input ports, the label state is a tensor product of the individual label states. In particular, for two particles,
$\hat{\varrho}^{(L)}=\hat{\rho}_1\otimes\hat{\rho}_2$.
Summing Eq.~(\ref{p{m,n}}) over all two-particle output occupations within the selected subset recovers Eq.~(\ref{pkl}).
The identity relative permutation $\pi\sigma^{-1}=I$ gives the direct term $H_{kk}H_{ll}$.
The transposition $\pi\sigma^{-1}=t$, where $t=(1,2)$ exchanges the two particles, gives the exchange term, with
$J(t)=\mathrm{Tr}(\hat{\rho}_1\hat{\rho}_2)$.
Its sign is determined by $\vare(t)$, which equals $1$ for bosons and $-1$ for fermions.

To  account for particle losses, we embed the lossy interferometer $U$ into a $2M\times2M$ unitary interferometer $\mathcal{U}$
\cite{Barnett1996,GenBunch2016}.
We choose its first $M$ rows as
$\mathcal{U}_{kj}=U_{kj}$ and
$\mathcal{U}_{k,M+j}=V_{kj}$, $k,j=1,\ldots,M$, with
$VV^\dag=I-UU^\dag$.
These rows are orthonormal and can therefore be completed to a unitary matrix.
The additional output modes represent losses, which are path-dependent, in general. The additional input modes are unoccupied, so the input occupation vector is $(\bm{n},\bm{0})$ (the additional input modes are not needed  below).

We apply Eq.~(\ref{p{m,n}}) to $\mathcal{U}$ with the extended output occupation vector
\[
(\bm{m},\bm{\ell})
=(m_1,\ldots,m_M,\ell_1,\ldots,\ell_M),
\qquad
|\bm{m}|+|\bm{\ell}|=n.
\]
The probability of detecting $n^\prime\le n$ particles with occupation vector $\bm{m}$, $|\bm{m}|=n^\prime$, is obtained by summing over all compatible occupations of the loss modes:
\be
p_{\bm{m},\bm{n}}(U):=
\sum_{|\bm{\ell}|=n-n^\prime}
p_{(\bm{m},\bm{\ell}),(\bm{n},\bm{0})}(\mathcal{U}).
\en{p{m,n}Loss}
The explicit expression is given in Ref.~\cite{ShchQuant2021} and is not needed in what follows.

\subsection{Marginal $r$-particle probabilities}
\label{Sec3C}

Let us now consider  the marginal output
probabilities~\cite{CollPhs2018,IndTh2026}.
For $2\le r\le n$, the $r$-particle marginal probability
associated with the output occupation vector
$\bm{s}=(s_1,\ldots,s_M)$,
$|\bm{s}|:=s_1+\ldots+s_M=r$, reads
\be
p^{(r|n)}_{\bm{s},\bm{n}}
=
\sum_{\bm{m}\supseteq\bm{s}}
P^{(\bm{m})}_{\bm{s}}p_{\bm{m},\bm{n}},
\quad
P^{(\bm{m})}_{\bm{s}}
:=
\binom{n}{r}^{-1}
\!\prod_{j=1}^M\binom{m_j}{s_j},
\en{pmarOUT}
where losses are included in $p_{\bm{m},\bm{n}}$ as described
above. The sum runs over all detected occupation vectors
$\bm{m}=(m_1,\ldots,m_M)$ satisfying
$m_j\ge s_j$ for every $j$ and
$r\le|\bm{m}|\le n$.
 
Equation~(\ref{pmarOUT}) can be understood by first considering
a lossless interferometer. The joint output distribution  is
symmetric under permutations of the interfering particles.
We retain $r$ output indices and sum over the remaining
indices $j_{r+1},\ldots,j_n$.
Grouping the terms by output occupations yields the
multivariate hypergeometric weight
$P^{(\bm{m})}_{\bm{s}}$.
For a lossy interferometer, we apply the same argument to
the lossless embedding described above, including the
unobserved loss modes in the sum over the remaining indices.
The weight $P^{(\bm{m})}_{\bm{s}}$ is then the probability
that a uniformly selected subset of $r$ of the $n$ particles
consists entirely of detected particles with occupation
vector $\bm{s}$, given the detected occupation vector
$\bm{m}$. 
 Accordingly,
\be
\sum_{|\bm{s}|=r}P^{(\bm{m})}_{\bm{s}}
=
\frac{\binom{|\bm{m}|}{r}}{\binom{n}{r}}
\le1.
\en{sumPr|n}

Equation~(\ref{pmarOUT}) allows the marginal probability
to be estimated directly from experimental data.
Its physical interpretation can be  made explicit by the 
following equivalent expression~\cite{CollPhs2018,IndTh2026}:
\be
p^{(r|n)}_{\bm{s},\bm{n}}
=
\sum_{\bm{n}^\prime}
P^{(\bm{n})}_{\bm{n}^\prime}
p_{\bm{s},\bm{n}^\prime},
\quad
P^{(\bm{n})}_{\bm{n}^\prime}
=
\binom{n}{r}^{-1}
\!\prod_{k=1}^M\binom{n_k}{n^\prime_k},
\en{pmarIN}
where the sum runs over input occupation vectors
$\bm{n}^\prime=(n^\prime_1,\ldots,n^\prime_M)$
satisfying $|\bm{n}^\prime|=r$ and
$0\le n^\prime_k\le n_k$.
The weights in Eq.~(\ref{pmarIN})  are given by the multivariate hypergeometric distribution
obtained by selecting $r$ particles uniformly from the
$n$ input particles.

The transition probability $p_{\bm{s},\bm{n}^\prime}$
is given by an expression analogous to
Eq.~(\ref{p{m,n}}), using the reduced label state of the
selected particles.
Their indistinguishability function is
\be
J_{\bm{n}^\prime}(\sigma)
=
\mathrm{Tr}\!\left[
\left(\hat{P}_\sigma\otimes\hat{I}^{\otimes(n-r)}\right)
\hat{P}_\nu^\dag\hat{\varrho}^{(L)}\hat{P}_\nu
\right],
\quad \sigma\in S_r,
\en{Jr|n}
where the permutation $\nu$ is chosen so that
$\nu(1),\ldots,\nu(r)$ label the selected particles
with input occupation vector $\bm{n}^\prime$.

\subsection{Marginal $r$-particle bunching}
\label{sec3D}

We now introduce the concept of marginal bunching: the probability that $r\ge 2$
particles selected uniformly from the $n$ input particles are
detected within a chosen proper subset of output ports.
It is the generalized bunching probability introduced in
Ref.~\cite{GenBunch2016}, averaged over all selections of
$r$ particles from the $n$ input particles.
Equivalently, the average runs over input occupation vectors
$\bm{n}^\prime$ satisfying $|\bm{n}^\prime|=r$ and
$0\le n_k^\prime\le n_k$, with the hypergeometric weights
$P^{(\bm{n})}_{\bm{n}^\prime}$ of Eq.~(\ref{pmarIN}).
Combining Eq.~(\ref{pmarIN}) with the expression for the generalized 
bunching probability $P(\bm{n}^\prime)$ of $r$ identical
particles~\cite{GenBunch2016}, we obtain
\begin{eqnarray}
\label{Prbunch}
&&P^{(r|n)}(\bm{n})
=\sum_{\bm{n}^\prime}
P^{(\bm{n})}_{\bm{n}^\prime}P(\bm{n}^\prime),\\
&&P(\bm{n}^\prime)
\equiv\frac{1}{\bm{n}^\prime!}
\sum_{\sigma\in S_r}\vare(\sigma)J_{\bm{n}^\prime}(\sigma)
\prod_{\alpha=1}^r H_{k_\alpha,k_{\sigma(\alpha)}},
\nonumber
\end{eqnarray}
where $k_1,\ldots,k_r$ correspond to the selected input occupation
vector $\bm{n}^\prime$, and $J_{\bm{n}^\prime}$ is given by
Eq.~(\ref{Jr|n}).
 
 The marginal bunching probability in Eq.~(\ref{Prbunch})
can also be expressed directly in terms of the total detected
occupation in the selected subset $\Omega_K$.
Summing the marginal probabilities in Eq.~(\ref{pmarOUT})
over all $\bm{s}=(s_1,\ldots,s_K,0,\ldots,0)$ with
$|\bm{s}|=r$, we obtain
\be
P^{(r|n)}(\bm{n})
=\binom{n}{r}^{-1}\sum_{\bm{m}}
 \binom{|\bm{m}|_K}{r}\,
p_{\bm{m},\bm{n}},
\quad
|\bm{m}|_K:=\sum_{j=1}^K m_j.
\en{PrbunchOUT}
The sum runs over all detected output occupation vectors
satisfying $r\le|\bm{m}|\le n$, with
$\binom{|\bm{m}|_K}{r}=0$ when \mbox{$|\bm{m}|_K<r$.}
Thus, an unbiased estimator of $P^{(r|n)}(\bm{n})$
 on $R$ runs is the sample mean of a bounded
observable \mbox{$0\le Z_i\le1$:}
\be
\widehat{P}^{(r|n)}_R
=
\frac{1}{R}\sum_{i=1}^R Z_i,
\qquad
Z_i:= \binom{n}{r}^{-1}\binom{N^{(i)}}{r},
\en{Zi}
where  $N^{(i)}\equiv |\bm{m}^{(i)}|_K$,  the number of particles
detected in the subset $\Omega_K$ in the $i$th run.
The sum includes all runs, assigning $Z_i=0$ whenever
$N^{(i)}<r$, without conditioning on particle survival.
Since $\mathbb{E}[Z_i]=P^{(r|n)}(\bm{n})$, we have
$\mathbb{E}[\widehat{P}^{(r|n)}_R]=P^{(r|n)}(\bm{n})$.

\section{Indistinguishability guarantees from two-particle marginal bunching}
\label{Sec4}

For $r=2$, Eq.~(\ref{Prbunch}) becomes
\begin{eqnarray}
\label{Pb2}
P^{(2|n)}(\bm{n})&=&
\sum_k\frac{n_k(n_k-1)}{n(n-1)}P(k,k)
\nonumber\\
&+&\sum_{k<l}\frac{2n_kn_l}{n(n-1)}P(k,l),\\
P(k,k)&=&H_{kk}^2,\quad
P(k,l)=H_{kk}H_{ll}\pm J(t_{kl})|H_{kl}|^2,\nonumber
\end{eqnarray}
where $t_{kl}$ exchanges two particles selected from distinct input ports $k$ and $l$, one from each port. Different choices of these particles are related by permutations in the subgroup appearing in Eq.~(\ref{JYoung}) and give the same value of $J(t_{kl})$.

Only pairs selected from distinct input ports contribute terms that depend on the indistinguishability function. Their probabilities have the form of Eq.~(\ref{pkl}), with the product-state overlap
$\mathrm{Tr}(\hat{\rho}_1\hat{\rho}_2)$ replaced by the general pairwise expectation $J(t_{kl})$.

\subsection{Indistinguishability guarantee}
\label{Sec4A}

Recall that $n$ identical particles are indistinguishable if and only if
$J(\sigma)=1$ for every $\sigma\in S_n$~\cite{IndTh2015}.
In this case, Eq.~(\ref{p{m,n}}) reduces to the squared modulus of a multiparticle transition amplitude (in case of bosons, divided by $\bm{m}!\bm{n}!$). This amplitude is a matrix permanent for bosons and a determinant for fermions.

We first show that indistinguishability of every particle pair implies multiparticle indistinguishability.

\begin{proposition}
\label{prop2}
The only indistinguishability functions satisfying $J(t)=\pm1$ for every transposition $t$, with the same sign for all transpositions, are
$J(\sigma)\equiv1$ and
$J(\sigma)\equiv\mathrm{sgn}(\sigma)$, respectively.
\end{proposition}

\begin{proof}
Let $s\in\{1,-1\}$ and suppose that $J(t)=s$ for every transposition.
Write the spectral decomposition of the label state as
\[
\hat{\varrho}^{(L)}
=\sum_i p_i|\phi_i\rangle\langle\phi_i|,
\qquad p_i>0,\quad \sum_i p_i=1.
\]
Since each $\hat{P}_t$ is Hermitian and unitary, Eq.~(\ref{J}) gives
\[
\sum_i p_i
\left\|(\hat I-s\hat P_t)|\phi_i\rangle\right\|^2
=2[1-sJ(t)]=0.
\]
Every term is nonnegative, so
$\hat P_t|\phi_i\rangle=s|\phi_i\rangle$
for every transposition $t$ and every $i$.
Every permutation is a product of transpositions. Thus, each vector in the support of $\hat{\varrho}^{(L)}$ is symmetric when $s=1$ and antisymmetric when $s=-1$.
Taking the expectation of $\hat P_\sigma$ therefore gives
$J(\sigma)=1$ or $J(\sigma)=\mathrm{sgn}(\sigma)$, respectively.
\end{proof}

The case $J(\sigma)\equiv\mathrm{sgn}(\sigma)$ corresponds to particles of one species emulating indistinguishable particles of the other species.
Bosons can emulate indistinguishable fermions only when the occupied input ports each contain a single particle.
Fermions can emulate indistinguishable bosons for arbitrary input occupations~\cite{BSF2015}, provided that the single-particle label Hilbert space has dimension at least $n$, so that an antisymmetric $n$-particle label state exists.

The sensitivity of Eq.~(\ref{Pb2}) to a given pairwise expectation $J(t_{kl})$ is determined by $|H_{kl}|^2$. We impose the condition that this sensitivity is nonzero for every pair of occupied input ports.

\begin{definition}[Generic setup]
Consider a linear interferometer with $M\ge2$ input and output ports and at least two occupied input ports.
The setup is called generic if there exists a nonempty proper subset of output ports for which the matrix $H$, defined in Eq.~(\ref{pkl}), satisfies
\[
\omega_{\min}:=\min_{k<l}|H_{kl}|^2>0,
\]
where the minimum is taken over pairs of occupied input ports.
Any output subset satisfying this condition is also called generic.
\end{definition}

To express the extremal probabilities, we introduce
\begin{eqnarray}
&&h_{kl}:=\frac{H_{kl}}{K}
=\frac{1}{K}\sum_{j=1}^K U_{kj}U^*_{lj},
\quad
p_1:=\sum_k\frac{n_k}{n}h_{kk},
\nonumber\\
&&c_1:=\sum_k\frac{n_k}{n}h_{kk}^2,
\quad
c_2:=2\sum_{k<l}\frac{n_kn_l}{n^2}|h_{kl}|^2.
\label{params}
\end{eqnarray}
Here $p_1$ is the single-particle transition probability averaged over input ports with weights $n_k/n$ and uniformly over the $K$ selected output ports.
The parameter $c_1$ is the weighted second moment of the output-port-averaged probabilities $h_{kk}$, while $c_2$ is a weighted sum of the squared off-diagonal elements $|h_{kl}|^2$.

Equation~(\ref{Pb2}) and Proposition~\ref{prop2} yield the following result.

\begin{theorem}[Marginal two-particle bunching]
\label{th1}
Consider $n\ge2$ identical particles in a generic setup.
For fermions, assume that the single-particle label Hilbert space has dimension at least $n$.
The two-particle marginal bunching probability in any generic subset of $K<M$ output ports attains its maximum
\be
P^{(2|n)}_{\mathrm{max}}
=
\frac{nK^2}{n-1}
\left(p_1^2-\frac{c_1}{n}+c_2\right)
\en{max2Bunch}
if and only if $J(\sigma)\equiv1$ for bosons or
$J(\sigma)\equiv\mathrm{sgn}(\sigma)$ for fermions.
\end{theorem}

\begin{proof}
For bosons, corresponding to the plus sign in Eq.~(\ref{Pb2}), the bound $J(t)\le1$ gives an upper bound on each pair contribution.
Genericity ensures that every coefficient multiplying $J(t_{kl})$ is positive, so equality requires
$J(t_{kl})=1$ for all pairs of distinct occupied input ports.
Similarly, for fermions, the minus sign in Eq.~(\ref{Pb2}) and the bound $J(t)\ge-1$ imply that equality requires
$J(t_{kl})=-1$ for all such pairs.

For transpositions within the same input port,
Eq.~(\ref{JYoung}) already requires $J(t)=1$ for bosons and $J(t)=-1$ for fermions.
Consequently, saturation requires $J(t)=\vare(t)$ for every transposition in $S_n$.
Proposition~\ref{prop2} gives the stated form of $J(\sigma)$.
The corresponding symmetric or antisymmetric label states exist under the assumptions of the theorem and attain the bound.
Substitution into Eq.~(\ref{Pb2}), followed by use of Eq.~(\ref{params}), gives Eq.~(\ref{max2Bunch}).
\end{proof}

For particles occupying distinct input ports, the minimum is characterized similarly.

\begin{corollary}[Marginal two-particle anti-bunching]
\label{col1}
Consider $n\ge2$ identical particles occupying distinct input ports in a generic setup.
For bosons, assume that the single-particle label Hilbert space has dimension at least $n$.
The two-particle marginal bunching probability in any generic subset of $K<M$ output ports attains its minimum
\be
P^{(2|n)}_{\mathrm{min}}
=
\frac{nK^2}{n-1}
\left(p_1^2-\frac{c_1}{n}-c_2\right)
\en{min2Bunch}
if and only if $J(\sigma)\equiv\mathrm{sgn}(\sigma)$ for bosons or
$J(\sigma)\equiv1$ for fermions.
\end{corollary}

Indeed, with distinct input ports, every particle pair contributes to Eq.~(\ref{Pb2}). The minimum requires $J(t)=-1$ for bosons or $J(t)=1$ for fermions for every transposition, and Proposition~\ref{prop2} applies.

Equation~(\ref{min2Bunch}) remains a lower bound when input ports are multiply occupied, but equality is impossible in a generic setup.
To see this, take particles $a$ and $b$ from the same input port and particle $c$ from another occupied port.
Equality would require the cross-port transpositions $(a,c)$ and $(b,c)$ to act on the support of the label state with the same sign, opposite to the within-port sign required by Eq.~(\ref{JYoung}).
However,
\[
(a,b)=(a,c)(b,c)(a,c)
\]
then forces the within-port transposition $(a,b)$ to act with that same opposite sign, giving a contradiction.

Thus, two-boson marginal bunching in a generic output subset is maximized only by indistinguishable bosons.
For fermions occupying distinct input ports, two-fermion marginal bunching is minimized only by indistinguishable fermions.

\subsection{Separable label states}
\label{Sec4B}

An important practical case is that of particles occupying distinct input ports with a separable label state, that is, a convex mixture of product states~\cite{IndTh2015,GenBunch2016},
\be
\hat{\varrho}^{(L)}_S=
\sum_i p_i\,
\hat{\rho}^{(i)}_1\otimes\ldots\otimes\hat{\rho}^{(i)}_n,
\qquad p_i\ge0,\quad \sum_i p_i=1.
\en{varrhoLS}
The indistinguishability function has the corresponding expansion
$J(\sigma)=\sum_i p_iJ^{(i)}(\sigma)$.
For a transposition $t=(a,b)$,
\[
J^{(i)}(t)
=\mathrm{Tr}\!\left(\hat{\rho}^{(i)}_a\hat{\rho}^{(i)}_b\right)
\ge0,
\]
and hence $J(t)\ge0$.

It follows from Eq.~(\ref{Pb2}) that the marginal bunching probability for distinguishable particles \cite{IndTh2015}, $J(\sigma) = \delta_{\sigma,I}$ (where $I$ denotes the identity permutation), which reads 
\be
P^{(2|n)}_{\mathrm{d}}
=
\frac{nK^2}{n-1}
\left(p_1^2-\frac{c_1}{n}\right),
\en{P2|n_Dist}
is a lower bound for bosons and an upper bound for fermions with separable label states.

For a generic output subset, equality holds if and only if $J(t)=0$ for every transposition.
Within the separable class, this condition is equivalent to
$J(\sigma)=\delta_{\sigma,I}$ for every $\sigma\in S_n$.
Indeed, the nonnegativity of every product-state contribution implies that $J(t)=0$ for all pairs only if the individual label states have mutually orthogonal supports in every component with $p_i>0$.
Every nonidentity permutation then has zero expectation in each component and hence in their mixture.

Whenever a distinguishable preparation is available, $P^{(2|n)}_{\mathrm{d}}$ is therefore the attainable minimum for bosons and maximum for fermions in  separable label states.
Such a preparation exists when the single-particle label Hilbert space has dimension at least $n$.

\section{Marginal three-boson bunching}
\label{Sec5}

The maximality property of Theorem~\ref{th1} extends to three-boson marginals. Using the results of Refs.~\cite{GenBunch2016,BapatSunder1986} and Proposition~\ref{prop2}, we obtain the following theorem, proved in Appendix~\ref{appE}.

\begin{theorem}[Marginal three-boson bunching]
\label{th2}
For $n\ge3$ identical bosons in a generic setup, the three-boson marginal bunching probability in any generic subset of $K<M$ output ports attains its maximum
\be
P^{(3|n)}_{\mathrm{max}}
=
\sum_{\substack{|\bm{n}^\prime|=3\\
0\le n^\prime_k\le n_k}}
P^{(\bm{n})}_{\bm{n}^\prime}
\frac{\mathrm{per}(H[\bm{n}^\prime])}{\bm{n}^\prime!}
\en{max3Bunch}
if and only if $J(\sigma)\equiv1$. Here $H[\bm{n}^\prime]$ is the $3\times3$ matrix with entries
\[
\bigl(H[\bm{n}^\prime]\bigr)_{\alpha\beta}
=H_{k_\alpha k_\beta},
\qquad \alpha,\beta=1,2,3,
\]
where $k_1\le k_2\le k_3$ encode the input occupation vector $\bm{n}^\prime$, with repeated indices when an input port is multiply occupied.
\end{theorem}

Three-particle marginals contain contributions from three-cycle expectations, which can carry collective-phase information not determined by pairwise indistinguishability~\cite{3Phas2017,CollPhs2018}. Theorem~\ref{th2} therefore provides a basis for future extending the certification method to incorporate this additional information.

\section{Indistinguishability certification by two-particle marginal bunching}
\label{Sec6}

The certification bounds in this section concern $n\ge2$ particles occupying distinct input ports, which we label by $k=1,\ldots,n$.
We choose a family of $q\ge1$ nonempty proper output subsets
$\Omega_\alpha$, $\alpha=1,\ldots,q$, with
$|\Omega_\alpha|=K_\alpha<M$.
The subsets may overlap.
We assign them positive weights $w_\alpha$ satisfying
$\sum_{\alpha=1}^q w_\alpha=1$, and define
\be
\omega_{kl}:=\sum_{\alpha=1}^q
w_\alpha|H^{(\alpha)}_{kl}|^2,
\qquad
H^{(\alpha)}_{kl}:=
\sum_{j\in\Omega_\alpha}U_{kj}U^*_{lj}.
\en{omegakl}

We now substitute  the genericity condition of Section~\ref{Sec4A}, applicable to a single output subset,  by the following one, applicable to   a family of   subsets.

\begin{definition}[Generic family of output subsets]
\label{def:generic_family}
For particles occupying distinct input ports, a family
$\Omega_1,\ldots,\Omega_q$ is generic if
$\omega_{kl}>0$ in Eq.~(\ref{omegakl}) for every pair of occupied input ports $k<l$.
\end{definition}

Definition~2 generalizes Definition~1 to families of output subsets: for every pair of occupied input ports $k<l$, at least one subset $\Omega_\alpha$ must satisfy $H^{(\alpha)}_{kl}\ne0$. The subset may depend on the pair, so a family can be generic even if none of its members is generic.

Consider the weighted two-particle marginal bunching probability
\be
P^{(2|n)}_T:=
\sum_{\alpha=1}^q w_\alpha P^{(2|n)}_{\Omega_\alpha},
\en{P_T}
where $P^{(2|n)}_{\Omega_\alpha}$ is given by Eq.~(\ref{Pb2}) with $H$ replaced by $H^{(\alpha)}$.
Its upper and lower bounds are
\be
P^{(2|n)}_{T\max}
=\sum_{\alpha=1}^q w_\alpha P^{(2|n)}_{\alpha,\max},
\qquad
P^{(2|n)}_{T\min}
=\sum_{\alpha=1}^q w_\alpha P^{(2|n)}_{\alpha,\min},
\en{Pwmaxmin}
where $P^{(2|n)}_{\alpha,\max}$ and
$P^{(2|n)}_{\alpha,\min}$ are obtained from
Eqs.~(\ref{max2Bunch}) and (\ref{min2Bunch}), using
$H^{(\alpha)}$ and $K_\alpha$ in Eq.~(\ref{params}).

For a generic family, the equality conditions of
Theorem~\ref{th1} and Corollary~\ref{col1} apply to these weighted bounds, with the same assumptions on the label Hilbert space.
Whenever the corresponding label state exists, it saturates all subset bounds simultaneously.
In particular, indistinguishable particles attain
$P^{(2|n)}_{T\max}$ for bosons and
$P^{(2|n)}_{T\min}$ for fermions.

Define the weighted bunching or anti-bunching probability defect by
\be
\Delta P^{(2|n)}_T
=
\begin{cases}
P^{(2|n)}_{T\max}-P^{(2|n)}_T,
&\textrm{bosons},\\[2pt]
P^{(2|n)}_T-P^{(2|n)}_{T\min},
&\textrm{fermions}.
\end{cases}
\en{DeltP}
Equation~(\ref{Pb2}) then gives
\begin{eqnarray}
\label{Deltaw}
&&\Delta P^{(2|n)}_T
=\binom{n}{2}^{-1}
\mathrm{Tr}\!\left(
\hat{\Delta}_\omega\hat{\varrho}^{(L)}
\right),\\
&&\hat{\Delta}_\omega:=
\sum_{k<l}\omega_{kl}
\left(\hat I-\hat P_{kl}\right),\nonumber
\end{eqnarray}
where $\hat P_{kl}:=\hat P_{t_{kl}}$ exchanges the internal tensor factors associated with particles entering ports $k$ and $l$.

The defect operator $\hat{\Delta}_\omega$ acts on
$\mathcal H_{\mathrm{int}}^{\otimes n}$ and is positive semidefinite, since each $\hat P_{kl}$ is Hermitian and unitary.
For a generic family, its kernel is the symmetric subspace, by Proposition~\ref{prop2}.
Writing the corresponding projector as
\[
\hat S_n:=\frac{1}{n!}\sum_{\sigma\in S_n}\hat P_\sigma,
\]
we obtain
\be
\hat{\Delta}_\omega
\ge
g(\hat{\Delta}_\omega)
\left(\hat I-\hat S_n\right),
\en{Deltagap}
where $g(\hat{\Delta}_\omega)$ denotes the smallest positive eigenvalue.
The zero eigenvalue generally has multiplicity
$\dim\operatorname{Sym}^n(\mathcal H_{\mathrm{int}})$.

We assume that $\dim\mathcal H_{\mathrm{int}}\ge2$.
For finite dimension $d$, the defect operator is represented by a matrix of size $d^n$.
Its spectral gap, however, can be obtained from an $n\times n$ matrix.
Let $T_{kl}$ exchange coordinates $k$ and $l$ in $\mathbb R^n$, and define, for a generic selection,  the following positive semidefinite symmetric matrix 
\be
L_\omega:=
\sum_{k<l}\omega_{kl}(I-T_{kl}),
\qquad
(L_\omega x)_k=
\sum_{l\ne k}\omega_{kl}(x_k-x_l).
\en{Lomega}
Its kernel is spanned by
$x^{(0)}=(1,\ldots,1)/\sqrt n$.
Thus,
\be
L_\omega\ge
g(L_\omega)\left(I-\frac{E}{n}\right),
\qquad E_{ij}=1,
\en{Lgap}
where $g(L_\omega)=\lambda_2(L_\omega)>0$.

To relate the two operators, choose orthonormal internal states
$|0\rangle$ and $|1\rangle$ and define the isometric embedding
$\mathcal M:\mathbb C^n\to\mathcal H_{\mathrm{int}}^{\otimes n}$ by
\[
\mathcal M e_k
=
|0\rangle^{\otimes(k-1)}
\otimes|1\rangle
\otimes|0\rangle^{\otimes(n-k)},
\]
where $e_k$ is the $k$th standard basis vector.
The identity
\[
\hat{\Delta}_\omega\mathcal M=\mathcal M L_\omega
\]
shows that $g(\hat{\Delta}_\omega)\le g(L_\omega)$.

Aldous' spectral gap theorem, proved by Caputo, Liggett, and Richthammer~\cite{SpecGap}, gives the reverse inequality: on every nontrivial irreducible representation of $S_n$, the smallest eigenvalue of the weighted transposition operator is at least $g(L_\omega)$.
Consequently,
\[
g(\hat{\Delta}_\omega)=g(L_\omega).
\]
The gap in Eq.~(\ref{Deltagap}) can therefore be computed by diagonalizing $L_\omega$, requiring   $\mathcal O(n^3)$ arithmetic operations.

We can now state the quantitative certification bound.

\begin{proposition}
\label{prop3}
For a generic family of output subsets, the projective indistinguishability measure~\cite{TightBndBS2015}
\be
\mathcal D:=
\frac{1}{n!}\sum_{\sigma\in S_n}J(\sigma)
=
\mathrm{Tr}\!\left(
\hat S_n\hat{\varrho}^{(L)}
\right)
\en{Dmeasure}
satisfies
\be
\mathcal D
\ge
1-\frac{
\mathrm{Tr}\!\left(
\hat{\Delta}_\omega\hat{\varrho}^{(L)}
\right)}
{g(L_\omega)}
=
1-\frac{
\binom{n}{2}\Delta P^{(2|n)}_T}
{g(L_\omega)}.
\en{Dboudgap}
For every generic family, the bound can be saturated at any value of $\mathcal D\in[0,1]$.
\end{proposition}

\begin{proof}
Taking the expectation of Eq.~(\ref{Deltagap}) and using
$g(\hat{\Delta}_\omega)=g(L_\omega)$ gives
Eq.~(\ref{Dboudgap}).

To establish tightness, let $x^{(g)}$ be a normalized eigenvector of $L_\omega$ with eigenvalue $g(L_\omega)$, and set
$|g\rangle:=\mathcal M x^{(g)}$.
Then
\[
\hat{\Delta}_\omega|g\rangle
=g(L_\omega)|g\rangle.
\]
Moreover, $x^{(g)}\perp x^{(0)}$ and
$\hat S_n\mathcal M x=\mathcal M(E/n)x$, so
$\hat S_n|g\rangle=0$.
For any normalized symmetric state $|\phi\rangle$ and
$0\le\epsilon\le1$, consider
\be
\hat{\varrho}^{(L)}_g
:=
(1-\epsilon)|\phi\rangle\langle\phi|
+\epsilon|g\rangle\langle g|.
\en{labelg}
This state satisfies
\[
\mathcal D(\hat{\varrho}^{(L)}_g)=1-\epsilon,
\qquad
\mathrm{Tr}\!\left(
\hat{\Delta}_\omega\hat{\varrho}^{(L)}_g
\right)
=g(L_\omega)\epsilon,
\]
and therefore saturates Eq.~(\ref{Dboudgap}).
\end{proof}

\begin{figure}[htb]
\includegraphics[width=\linewidth]{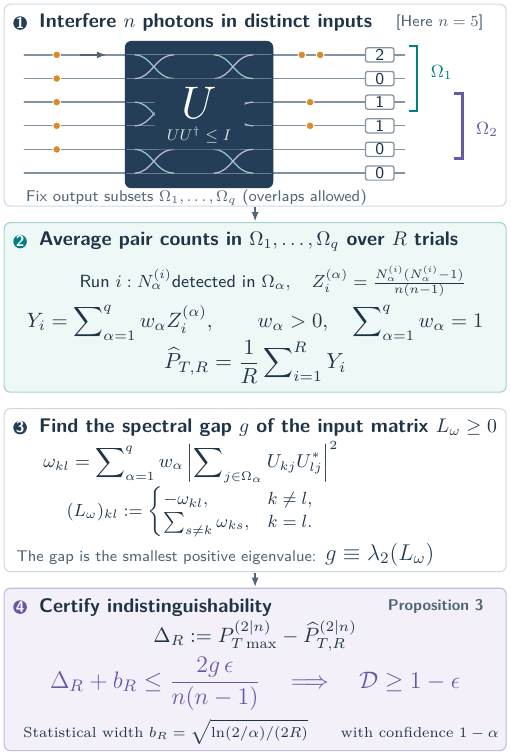}
\caption{\textit{In situ} certification of $n$-photon indistinguishability using Proposition~\ref{prop3}.
(1) Photons occupying distinct input ports interfere in a possibly lossy interferometer $U$.
(2) Number-resolved output records yield a weighted pair count for fixed, possibly overlapping output subsets, with
$w_\alpha>0$ and $\sum_\alpha w_\alpha=1$.
All trials, including loss events, contribute to the sample mean.
(3) Construct the matrix  $L_\omega$ on the occupied input ports and determine its spectral gap
$g=\lambda_2(L_\omega)$.
 (4) Comparing the measured mean with the ideal bosonic maximum
$P^{(2|n)}_{T\max}$ gives a lower bound on $\mathcal D$ with confidence at least $1-\alpha$.
Here $b_R$ is the Hoeffding error margin for $R$ independent trials.
The schematic shows five input photons in a six-mode interferometer, with one photon lost and two overlapping output subsets.}
\label{fig1}
\end{figure}

Numerical diagonalization is generally needed to obtain
$g(L_\omega)$.
Alternatively, the following explicit lower bound gives a valid, generally weaker certification bound:
\be
g(L_\omega)\ge b_\omega:=
\min_{k<l}
\left[
\sum_{m\ne k,l}
\min\{\omega_{mk},\omega_{ml}\}
+2\omega_{kl}
\right].
\en{Lboundg}
To prove this inequality, take the (real)  eigenvector $x^{(g)}$ for $g(L_\omega)$. Let the indices $k$ and $l$  refer to its   components attaining    maximum and minimum.
Subtracting the corresponding eigenvalue equations gives
\begin{eqnarray*}
&&g(L_\omega)\bigl(x^{(g)}_k-x^{(g)}_l\bigr)\\
&&\quad=
\sum_{m\ne k,l}
\left[
\omega_{mk}\bigl(x^{(g)}_k-x^{(g)}_m\bigr)
+\omega_{ml}\bigl(x^{(g)}_m-x^{(g)}_l\bigr)
\right]\\
&&\qquad
+2\omega_{kl}\bigl(x^{(g)}_k-x^{(g)}_l\bigr).
\end{eqnarray*}
For each $m$, the two differences in the summand are nonnegative and add to
$x^{(g)}_k-x^{(g)}_l$.
The summand is therefore at least
\[
\min\{\omega_{mk},\omega_{ml}\}
\bigl(x^{(g)}_k-x^{(g)}_l\bigr).
\]
Since $x^{(g)}$ is nonconstant,
$x^{(g)}_k-x^{(g)}_l>0$, and division gives
Eq.~(\ref{Lboundg}).

Each term in Eq.~(\ref{Lboundg}) is bounded below by the smallest pair weight, yielding the coarser bound
\be
g(L_\omega)\ge n\min_{k<l}\omega_{kl}.
\en{glowb}
For uniform   $\omega_{kl}=\omega$, both bounds are exact:
\[
L_\omega=n\omega\left(I-\frac{E}{n}\right),
\qquad
g(L_\omega)=b_\omega=n\omega.
\]
In particular, the normalization $\omega=1/n$ gives unit spectral gap.

\subsection{Separable label states}
\label{Sec6A}

Proposition~\ref{prop3} is tight for every generic family, although the saturating label state can depend on the pair weights.
For separable label states, Eq.~(\ref{varrhoLS}), we can also compare the bounds obtained from different choices of weights.

\begin{corollary}
\label{col2}
For a fixed separable label state, the bound in Proposition~\ref{prop3} obtained with uniform $\omega_{kl}=\omega>0$  is at least as strong as that obtained with any other  generic $\omega_{kl}$.
Thus, whenever uniform $\omega>0$  can be realized through Eq.~(\ref{omegakl}), they give the strongest lower bound within this family.
\end{corollary}

\begin{proof}
For a product label state
$\hat{\varrho}^{(L)}=\bigotimes_{k=1}^n\hat\rho_k$,
define
\[
G_{kl}
=
\mathrm{Tr}(\hat\rho_k\hat\rho_l)
+\delta_{kl}
\left[1-\mathrm{Tr}(\hat\rho_k^2)\right].
\]
The first term is a positive semidefinite Gram matrix, and the second is a diagonal matrix with nonnegative entries.
Hence $G$ is positive semidefinite.

For the separable mixture in Eq.~(\ref{varrhoLS}), take the corresponding convex combination of these matrices.
The resulting $G$ is again positive semidefinite and satisfies
$G_{kk}=1$ and $G_{kl}=J(t_{kl})$ for $k\ne l$.

Introduce the uniformly weighted defect operator
\[
\hat{\Delta}_{1/n}
:=
\frac1n\sum_{k<l}
\left(\hat I-\hat P_{kl}\right).
\]
Using Eq.~(\ref{Lgap}) and the positivity of $G$, we obtain
\begin{eqnarray}
\label{Deltaw_S}
&&
\frac{
\mathrm{Tr}\!\left(
\hat{\Delta}_\omega\hat{\varrho}^{(L)}_S
\right)}
{g(L_\omega)}
=
\frac{\mathrm{Tr}(L_\omega G)}{2g(L_\omega)}
\nonumber\\
&&\quad\ge
\frac12\mathrm{Tr}\!\left[
\left(I-\frac En\right)G
\right]
=
\mathrm{Tr}\!\left(
\hat{\Delta}_{1/n}\hat{\varrho}^{(L)}_S
\right).
\end{eqnarray}
Here $g(L_{1/n})=1$, and
\be
\mathrm{Tr}\!\left(
\hat{\Delta}_{1/n}\hat{\varrho}^{(L)}_S
\right)
=
\frac1n\sum_{k<l}\left[1-J(t_{kl})\right].
\en{Delta1/n}
For any uniform  $\omega>0$,
$\hat{\Delta}_\omega=n\omega\hat{\Delta}_{1/n}$ and
$g(L_\omega)=n\omega$.
Thus, the right-hand side of Eq.~(\ref{Deltaw_S}) is the normalized defect entering the uniform-$\omega$ bound, which proves the comparison.

The uniform-$\omega$  bound can also be saturated by separable states arbitrarily close to $\mathcal D=1$.
Indeed, consider
\[
|\Psi_\epsilon\rangle
=
|0\rangle^{\otimes(n-1)}
\otimes
\left(
\sqrt{1-\epsilon}\,|0\rangle
+\sqrt{\epsilon}\,|1\rangle
\right),
\qquad 0\le\epsilon\le1.
\]
For this product state,
\[
\mathcal D(|\Psi_\epsilon\rangle)
=1-\frac{n-1}{n}\epsilon,
\qquad
\langle\Psi_\epsilon|
\hat{\Delta}_{1/n}
|\Psi_\epsilon\rangle
=\frac{n-1}{n}\epsilon,
\]
so equality holds in Eq.~(\ref{Dboudgap}).
\end{proof}

The uniform-$\omega$ bound can be expressed through the two-particle marginal projective measure of indistinguishability~\cite{IndTh2026},
\be
\mathcal D^{(2|n)}
=
\frac12
\left[
1+\binom n2^{-1}
\sum_{k<l}J(t_{kl})
\right],
\en{D2|n}
which is the average of the projective measures for all particle pairs.
Since
\be
\frac1n\sum_{k<l}\left[1-J(t_{kl})\right]
=
(n-1)\left(1-\mathcal D^{(2|n)}\right),
\en{D2nJ}
Proposition~\ref{prop3} with uniform$\omega_{kl}=\omega$ reduces to the bound of Ref.~\cite{NewDist2024},
\[
\mathcal D
\ge
1-(n-1)\left(1-\mathcal D^{(2|n)}\right).
\]
This inequality holds for arbitrary label states; separability  makes it optimal   in Corollary~\ref{col2} in  comparison with  non-uniform  $\omega_{kl}$.

Uniform $\omega_{kl}= \omega$ also allow a direct readout of
$\mathcal D^{(2|n)}$ from the weighted probability defect.
For example, suppose that one output port, chosen as $j=1$, is balanced over the occupied inputs:
\[
|U_{k1}|^2=X>0,
\qquad k=1,\ldots,n.
\]
Selecting this port alone gives $q=1$, $w_1=1$, and
$H_{kl}=U_{k1}U^*_{l1}$, so that
$\omega_{kl}=X^2=\omega$ for every pair.
More generally, whenever Eq.~(\ref{omegakl}) yields uniform weights, Eqs.~(\ref{Deltaw}) and (\ref{D2nJ}) give
\be
\Delta P^{(2|n)}_T
=
2\omega\left(1-\mathcal D^{(2|n)}\right).
\en{P2bX}
This readout also applies to arbitrary label states.

\subsection{Scaling of $P^{(2|n)}$ and certification cost}
\label{Sec6B}

We first record the Haar average of the bunching probability for a single output subset.
For this calculation, the input occupation vector may be arbitrary.
Keeping the input label state and the selected output subset fixed, integration over 
unitary interferometers of size $M$ with respect to the normalized Haar measure gives
(see Appendix~\ref{appA})
\begin{eqnarray}
\int dU\,P^{(2|n)}(\bm n)
&=&
\frac{K(K+1)}{M(M+1)}
\biggl(
1-\frac{M-K}{(M-1)(K+1)}
\nonumber\\
&&\quad\times
\sum_{k<l}\frac{2n_kn_l}{n(n-1)}
\left[1\mp J(t_{kl})\right]
\biggr).
\label{avP2b}
\end{eqnarray}
The prefactor has order $(K/M)^2$.
The fraction of particle pairs occupying distinct input ports is
\[
\beta:=
\frac{2\sum_{k<l}n_kn_l}{n(n-1)}.
\]
The combined coefficient of the pair-dependent terms is therefore
$\beta(M-K)/[(M-1)(K+1)]$.
For $M\gg1$, choosing $K=O(1)$ retains an order-one relative sensitivity to pairwise indistinguishability whenever
$\beta=\Theta(1)$.
For particles occupying distinct input ports, as required for the certification bounds, $\beta=1$.

For fixed output subsets and weights, the weighted probability is estimated from $R$ runs by
\[
\widehat P_T^{(2|n)}(R)
=
\frac1R\sum_{i=1}^R
\sum_{\alpha=1}^q
w_\alpha
\frac{
N_\alpha^{(i)}(N_\alpha^{(i)}-1)}
{n(n-1)},
\]
where $N_\alpha^{(i)}$ is the number of particles detected in
$\Omega_\alpha$ in run $i$.
All runs are included.
Each run contributes a value in $[0,1]$, so Hoeffding's inequality applies when the runs are independent.
Overlapping subsets  may   yield    correlated  counts within a run.

For an indistinguishable input,  treating the interferometer parameters as known, the sufficient run bound
\be
R\ge
\frac12
\left[
\frac{n(n-1)}
{2g(L_\omega)\epsilon}
\right]^2
\ln\left(\frac2\alpha\right)
\en{Runs}
guarantees certification of $\mathcal D\ge1-\epsilon$ with probability at least $1-\alpha$, where
$0<\epsilon,\alpha<1$.
The confidence bound and the derivation of Eq.~(\ref{Runs}) are given in Appendix~\ref{appC}.

For a Haar-random unitary interferometer, this sufficient number of runs is polynomial in $(n,M)$ with high probability at fixed precision and confidence.
Specifically, take a single nonempty proper output subset
($q=1$, $w_1=1$), with both the occupied input ports and the output subset fixed independently of $U$.
For any fixed $c>0$, Appendix~\ref{appC} gives
\[
\frac{n}{g(L_\omega)}
\le
\frac{1}{\min_{k<l}\omega_{kl}}
\le
(nM)^{2c+2}
\]
with probability at least $1-(nM)^{-c}$ over $U$.

Two-particle marginal bunching also improves the trade-off between visibility and sampling cost compared with  full $n$-particle generalized bunching~\cite{GenBunch2016}.
For particles occupying distinct input ports, selecting
$K=O(1)$ preserves a nonvanishing contrast between the Haar-averaged probabilities for indistinguishable and distinguishable bosons, while the certification cost is polynomial in $(n,M)$ with high probability over the interferometer.
Within this Haar-averaged comparison, in the regime
$M=\Omega(n^2)$ with $M$ polynomial in $n$, retaining the same nonvanishing contrast through the  full $n$-particle bunching requires exponentially rare events and hence exponentially many runs.
This conclusion holds even if each method chooses its own output-subset size; see Appendix~\ref{appD}.

Uniform losses further favor marginal bunching.
A single-particle survival probability $0<\eta\le1$ multiplies the two-particle marginal bunching probability by $\eta^2$, whereas the full $n$-particle bunching probability is multiplied by $\eta^n$.
Under the same uniform loss, each $H^{(\alpha)}$ scales by $\eta$, so both $\Delta P_T^{(2|n)}$ and $g(L_\omega)$ scale by $\eta^2$.
Their ratio in Eq.~(\ref{Dboudgap}) is therefore unchanged, while the sufficient run bound in Eq.~(\ref{Runs}) increases by a factor $\eta^{-4}$.
Thus, polynomial sampling cost is preserved whenever the inverse survival probability is polynomially bounded.

\section{Conclusion}
\label{Sec7}

We have established two main results. First, we have introduced marginal $r$-particle bunching for $2\le r\le n$, with generalized $n$-particle bunching recovered for $r=n$. In generic setups, two-boson marginal bunching is maximized only by indistinguishable bosons, while two-fermion marginal bunching is minimized only by indistinguishable fermions occupying distinct input ports. The corresponding extremal probabilities are given explicitly. Related extremal results characterize the ideal emulation of one species by the other through appropriately entangled label states. We have also proved that three-boson marginal bunching is maximized only by indistinguishable bosons.

Second, for particles occupying distinct input ports, we have used Aldous' spectral gap theorem to derive a tight quantitative bound on multiparticle indistinguishability from the two-particle marginal bunching. This bound provides an \textit{in situ} certification method for generic linear interferometers. For Haar-random unitary interferometers, the number of experimental runs required to estimate the bound to precision $\epsilon$, at fixed confidence, is polynomial in $(n,M,1/\epsilon)$ with high probability over the choice of interferometer. Here $n$ is the particle number and $M$ is the interferometer size. Under uniform losses, polynomial sampling cost is preserved whenever the inverse single-particle survival probability is polynomially bounded.

Several questions remain open. Collective phases~\cite{3Phas2017,CollPhs2018,4Phas2020} arise in interference involving three or more partially distinguishable particles. Theorem~\ref{th2} provides a starting point for incorporating the three-cycle information accessible through three-boson marginals into quantitative certification. A natural question is whether this additional information can improve the efficiency or robustness of indistinguishability assessment.

Our results also provide a perspective on the physical significance of conjectures in multilinear algebra (cf.~Ref.~\cite{Pioge2026}). Although these conjectures fail in general, their validity at particular matrix sizes can have useful physical consequences through the concept of marginal bunching. For example, the Soules (a.k.a. permanent-on-top) conjecture,  associated with generalized bunching~\cite{GenBunch2016}, has been proved for $r=3$~\cite{BapatSunder1986}. A proof for $r=4$ has recently been announced~\cite{Lavi2026}, while counterexamples exist for $r=5$~\cite{POT2016,tran2022}. The Bapat--Sunder conjecture, similarly related to generalized bunching with separable label states~\cite{GenBunch2016}, fails for $r=7$~\cite{Drury2016}. These connections motivate determining the matrix sizes for which such inequalities hold and identifying the equality conditions required for indistinguishability certification.

\begin{acknowledgments}
 This work was   supported by  the National Council for Scientific and Technological Development (Conselho Nacional de Desenvolvimento Cient\'ifico e Tecnol\'ogico) of Brazil,  grant number 307507/2023-8.    AI tools have been  employed for checking technical details   and for   proofreading. 
 \end{acknowledgments}

  \appendix
 \onecolumngrid
 
 \section{Simplicity of the permanent eigenvalue for three particles}
\label{appE}

For a $3\times3$ matrix $H$, let $\widetilde H$ denote its Schur
power matrix. This $6\times6$ matrix is indexed by
$\sigma,\tau\in S_3$, with entries
\begin{equation}
\widetilde H_{\sigma,\tau}
=
\prod_{i=1}^{3}H_{\tau(i),\sigma(i)}.
\label{eq:simple-permanent-schur-power}
\end{equation}
Theorem~2 of Ref.~\cite{BapatSunder1986} establishes that
$\operatorname{per}(H)$ is the largest eigenvalue of
$\widetilde H$ whenever $H$ is Hermitian and positive
semidefinite. We prove that this eigenvalue is simple under
the nonvanishing condition below, and then apply the result
to prove Theorem~\ref{th2} of the main text.

\begin{proposition}
\label{prop4}
Let $H$ be a $3\times3$ Hermitian positive semidefinite matrix
satisfying $|H_{12}H_{23}H_{31}|>0$.
Then $\operatorname{per}(H)$ is a simple largest eigenvalue
of its Schur power matrix $\widetilde H$.
\end{proposition}

\begin{proof}
The assumptions imply $H_{ii}>0$ for $i=1,2,3$.
As in Ref.~\cite{BapatSunder1986}, choose an invertible
diagonal matrix $D$ that normalizes the diagonal entries
and makes the off-diagonal entries in the first row real
and positive. Thus,
\begin{equation}
A=D^\dagger H D
=
\begin{pmatrix}
1&d&e\\
d&1&f\\
e&\bar f&1
\end{pmatrix},
\qquad
d,e>0,\quad f=a+ib\ne0,
\label{eq:simple-permanent-normalization}
\end{equation}
where $a,b\in\mathbb R$.
Writing $c:=|\det D|^2>0$, we have
\begin{equation}
\widetilde A=c\widetilde H,
\qquad
\operatorname{per}(A)=c\operatorname{per}(H).
\label{eq:simple-permanent-scaling}
\end{equation}
It therefore suffices to prove the assertion for $A$.

Following Ref.~\cite{BapatSunder1986}, define
\begin{equation}
\widehat A_{ij}:=a_{ij}\operatorname{per}A(i,j),
\qquad
L=(\operatorname{per}A)I_3-\widehat A,
\label{eq:simple-permanent-natural-block}
\end{equation}
where $a_{ij}$ are the entries of $A$ and $A(i,j)$ is obtained
by deleting row $i$ and column $j$.
The permanent expansion along each row gives
$L\mathbf1_3=0$, where $\mathbf1_3=(1,1,1)^T$.

The principal-minor calculation in
Ref.~\cite{BapatSunder1986} establishes that $L$ is positive
semidefinite. We recall the calculation needed for strictness.
Let $\mathcal M$ denote the principal minor of $L$
corresponding to rows and columns $1,2$. Direct calculation gives
\be
\mathcal M
=
3d^2e^2a^2
+
(2dea+d^2+e^2)b^2
+
x^TCx,
\en{minorM}
where
\begin{equation}
C:=
\begin{pmatrix}
1-b^2&d&e\\
d&1&a\\
e&a&1
\end{pmatrix},
\qquad
x:=
\begin{pmatrix}
de\\da\\ea
\end{pmatrix}.
\label{eq:simple-permanent-auxiliary-matrix}
\end{equation}
The positivity of $C$ can be checked directly, including
when $A$ is singular. Indeed, $A\ge0$ implies
$d,e,|f|\le1$ and
\[
\det A
=
(1-d^2)(1-e^2)-(a-de)^2-b^2
\ge0.
\]
Consequently, $b^2\le(1-d^2)(1-e^2)$, and hence
$1-b^2-d^2\ge0$ and $1-b^2-e^2\ge0$.
Together with $|a|\le1$ and
\[
\det C=\det A+a^2b^2\ge0,
\]
these inequalities show that every principal minor of $C$
is nonnegative. Thus $C\ge0$.

Moreover, $|a|\le|f|\le1$ gives
\begin{equation}
2dea+d^2+e^2
\ge
d^2+e^2-2de
=
(d-e)^2
\ge0.
\label{eq:simple-permanent-coefficient}
\end{equation}
Every term in Eq.~\eqref{minorM} is therefore nonnegative.
For completeness, the diagonal entries of $L$ also satisfy
\[
\begin{aligned}
L_{11}&=d^2+e^2+2dea\ge(d-e)^2,\\
L_{22}&=d^2+|f|^2+2dea\ge(d-|f|)^2,\\
L_{33}&=e^2+|f|^2+2dea\ge(e-|f|)^2.
\end{aligned}
\]
Since $L\mathbf1_3=0$ and $L$ is Hermitian, its three
principal minors of order two are equal to $\mathcal M$,
and $\det L=0$. These observations also verify $L\ge0$ directly.

Under the present assumptions, $\mathcal M$ is strictly
positive. If $a\ne0$, then $3d^2e^2a^2>0$. If $a=0$,
then $f\ne0$ implies $b\ne0$, and the second term in
Eq.~\eqref{minorM} becomes $(d^2+e^2)b^2>0$.
Thus $\mathcal M>0$. Together with $L\mathbf1_3=0$,
this implies
\begin{equation}
\operatorname{rank}L=2,
\qquad
\ker L=\operatorname{span}\{\mathbf1_3\}.
\label{eq:simple-permanent-natural-kernel}
\end{equation}
Because $L\ge0$, both eigenvalues of $\widehat A$ on
$\mathbf1_3^\perp$ are strictly smaller than
$\operatorname{per}A$.

It remains to pass from $\widehat A$ to $\widetilde A$.
In the notation of Ref.~\cite{BapatSunder1986}, the natural
representation of $S_3$ decomposes as $\pi_0\oplus\pi$,
where $\pi_0$ is the trivial representation,
$\sigma\mapsto1$, and $\pi$ is the two-dimensional standard
representation on
\[
\mathbf1_3^\perp
=
\{(x_1,x_2,x_3)\in\mathbb C^3:
x_1+x_2+x_3=0\}.
\]
The regular representation decomposes as
$\pi_0\oplus\pi_{\mathrm{sgn}}\oplus\pi\oplus\pi$,
where $\pi_{\mathrm{sgn}}$ is the sign representation,
$\sigma\mapsto\operatorname{sgn}(\sigma)$.
Accordingly, the spectrum of $\widetilde A$ consists of
$\operatorname{per}(A)$, $\det(A)$, and the two eigenvalues
of $\widehat A$ on $\mathbf1_3^\perp$, each of the latter
occurring twice.
The eigenvalue in the sign sector is also strictly smaller
than $\operatorname{per}A$, since
\begin{equation}
\operatorname{per}A-\det A
=
2\bigl(d^2+e^2+|f|^2\bigr)>0.
\label{eq:simple-permanent-sign-gap}
\end{equation}
Hence $\operatorname{per}A$ is a simple largest eigenvalue
of $\widetilde A$.
Equation~\eqref{eq:simple-permanent-scaling} proves the
assertion for $H$.
The argument also applies when $H$ is singular.
\end{proof}

To complete the proof of Theorem~\ref{th2} of the main text, apply
Proposition~\ref{prop4} to each matrix $H[\bm n^\prime]$ associated
with a feasible three-particle input occupation vector.
In a generic setup, all its off-diagonal entries are nonzero,
including entries associated with repeated input indices,
for which the corresponding $H_{kk}$ is positive.
The Schur-power representation of the bunching
probability~\cite{GenBunch2016} then implies that
\[
P(\bm n^\prime)
\le
\frac{\operatorname{per}H[\bm n^\prime]}{\bm n^\prime!},
\]
with equality only if $J_{\bm n^\prime}(\sigma)\equiv1$
on $S_3$.

All feasible input occupation vectors have positive weights
in the marginal average in Eq.~\eqref{Prbunch}. Equality
in the marginal bound therefore requires equality for every
such three-particle input.
Every pair of particles belongs to a three-particle subset,
so this forces $J(t)=1$ for every transposition $t\in S_n$.
Proposition~\ref{prop2} then gives $J(\sigma)\equiv1$ on $S_n$.
Conversely, this indistinguishability function attains the
stated bound, completing the proof.

\section{Haar average of the marginal two-particle bunching probability}
\label{appA}

We average over the normalized Haar measure on the group $ U(M)$ of $M\times M$ unitary matrices, keeping the input occupation vector, the label state, and the chosen subset of $K$ output ports fixed. The unitary integration formula reads
\be
\int dU \prod_{\alpha=1}^r
U_{k_\alpha,j_\alpha}U^*_{k^\prime_\alpha,j^\prime_\alpha}
=
\sum_{\nu,\tau\in S_r}W_M(\nu\tau^{-1})
\prod_{\alpha=1}^r
\delta_{k^\prime_\alpha,k_{\nu(\alpha)}}
\delta_{j^\prime_\alpha,j_{\tau(\alpha)}}.
\en{avUU*}
Here $W_M$ is the unitary Weingarten function on $S_r$~\cite{collins2006}. It depends only on the cycle structure of its argument, i.e., on the lengths of the disjoint cycles in the permutation.  

For the products of matrix elements of $H$ appearing in the $r$-particle bunching probability in Eq.~(\ref{Prbunch}), we obtain
\be
\begin{aligned}
&\int dU\prod_{\alpha=1}^r H_{k_\alpha,k_{\sigma(\alpha)}}\\
&\quad =
\sum_{j_1,\ldots,j_r=1}^K
\int dU\prod_{\alpha=1}^r
U_{k_\alpha,j_\alpha}U^*_{k_{\sigma(\alpha)},j_\alpha}\\
&\quad =
\sum_{\nu,\tau\in S_r}W_M(\nu\tau^{-1})
\left(\prod_{\alpha=1}^r
\delta_{k_{\sigma(\alpha)},k_{\nu(\alpha)}}\right)
\left(\sum_{j_1,\ldots,j_r=1}^K
\prod_{\alpha=1}^r\delta_{j_\alpha,j_{\tau(\alpha)}}\right)\\
&\quad =
\sum_{\mu\in Y_{\bm{n}^\prime}}\sum_{\tau\in S_r}
K^{c(\tau)}W_M(\mu\sigma\tau).
\end{aligned}
\en{AvH}
Here $Y_{\bm{n}^\prime}:=S_{n^\prime_1}\times\cdots\times S_{n^\prime_M}$ is the Young subgroup associated with the input occupation vector $\bm{n}^\prime$, and $c(\tau)$ is the number of disjoint cycles of $\tau$, including fixed points. We have used the identities
\[
\prod_{\alpha=1}^r
\delta_{k_{\sigma(\alpha)},k_{\nu(\alpha)}}
=
\sum_{\mu\in Y_{\bm{n}^\prime}}\delta_{\nu,\mu\sigma},
\qquad
\sum_{j_1,\ldots,j_r=1}^K
\prod_{\alpha=1}^r\delta_{j_\alpha,j_{\tau(\alpha)}}
=
K^{c(\tau)}.
\]
The second identity follows because the output indices must be equal within each cycle of $\tau$, leaving $K$ independent choices per cycle. We  have renamed     $\tau^{-1}\to \tau$, since $c(\tau^{-1})=c(\tau)$.

For general $r\ge 2$, substituting Eq.~(\ref{AvH}) into Eq.~(\ref{Prbunch}) gives
\be
\begin{aligned}
\int dU P(\bm{n}^\prime)
&=
\frac{1}{\bm{n}^\prime!}
\sum_{\sigma\in S_r}\vare(\sigma)J_{\bm{n}^\prime}(\sigma)
\int dU\prod_{\alpha=1}^r H_{k_\alpha,k_{\sigma(\alpha)}}\\
&=
\sum_{\pi\in S_r}\vare(\pi)J_{\bm{n}^\prime}(\pi)
\sum_{\tau\in S_r}K^{c(\tau)}W_M(\pi\tau).
\end{aligned}
\en{AvPrbuch}
To obtain the second line, we set $\pi=\mu\sigma$ and use the Young-subgroup symmetry in Eq.~(\ref{JYoung}),
\be
J_{\bm{n}^\prime}(\sigma)
=
\vare(\mu)J_{\bm{n}^\prime}(\mu\sigma),
\qquad
\mu\in Y_{\bm{n}^\prime},\quad \sigma\in S_r.
\en{symJYoung}
Together with $\vare(\mu)\vare(\sigma)=\vare(\pi)$, this gives
$\vare(\sigma)J_{\bm{n}^\prime}(\sigma)
=\vare(\pi)J_{\bm{n}^\prime}(\pi)$.
The sum over $\mu$ then contributes the factor
$|Y_{\bm{n}^\prime}|=\bm{n}^\prime!$, which cancels the prefactor.

For the case of interest, $r=2$, the required averages are:
\be
\begin{aligned}
\int dU\,H_{kk}H_{ll}
&=
\delta_{kl}\frac{K(K+1)}{M(M+1)}
+(1-\delta_{kl})\frac{K(KM-1)}{M(M^2-1)},\\
\int dU\,|H_{kl}|^2
&=
\frac{K(M-K)}{M(M^2-1)},
\qquad k\ne l.
\end{aligned}
\en{AvP2bunch}
These follow from the known values of the Weingarten function for $M\ge 2$~\cite{collins2006},
\[
W_M(I)=\frac{1}{M^2-1},
\qquad
W_M(t)=-\frac{1}{M(M^2-1)},
\]
where $I$ is the identity permutation and $t$ is the transposition in $S_2$. In particular,
\[
\sum_{\tau\in S_2}W_M(\tau)=\frac{1}{M(M+1)}.
\]

We next average the exchange factor
$\vare(t)J_{\bm{n}^\prime}(t)$ over the multivariate hypergeometric distribution
$P^{(\bm{n})}_{\bm{n}^\prime}$. Since $\vare(t)=\pm1$ for bosons and fermions, respectively, we find
\be
\begin{aligned}
&\pm\sum_{\bm{n}^\prime}
P^{(\bm{n})}_{\bm{n}^\prime}J_{\bm{n}^\prime}(t)\\
&\quad =
\pm\binom{n}{2}^{-1}
\sum_{\bm{n}^\prime}
\left[\prod_{k=1}^M\binom{n_k}{n^\prime_k}\right]
J_{\bm{n}^\prime}(t)\\
&\quad =
\binom{n}{2}^{-1}
\left[
\sum_{k=1}^M\binom{n_k}{2}
\pm\sum_{k<l}n_kn_l J(t_{kl})
\right]\\
&\quad =
1-\sum_{k<l}\frac{2n_kn_l}{n(n-1)}
\left[1\mp J(t_{kl})\right].
\end{aligned}
\en{avJ2}
Here the sums over $\bm{n}^\prime$ run over all two-particle input occupation vectors satisfying $\bm{n}^\prime\subseteq\bm{n}$.
For two particles selected from the same input port, Eq.~(\ref{symJYoung}) gives $J_{\bm{n}^\prime}(t)=\vare(t)$, so their exchange factor is unity. The last equality follows from
\[
\sum_{k=1}^M\binom{n_k}{2}+\sum_{k<l}n_kn_l
=
\binom{n}{2}.
\]
Combining Eqs.~(\ref{AvP2bunch}) and (\ref{avJ2}) with Eq.~(\ref{Pb2}), we obtain
\be
\int dU P^{(2|n)}(\bm{n})
=
\frac{K(K+1)}{M(M+1)}
\left[
1-\frac{M-K}{(M-1)(K+1)}
\sum_{k<l}\frac{2n_kn_l}{n(n-1)}
\left[1\mp J(t_{kl})\right]
\right].
\en{avP2nbuch}

\section{Estimate of the number of experimental runs}
\label{appC}

We consider $n\ge2$ particles occupying distinct input ports, as required by the certification bounds in Section~\ref{Sec6}.
The interferometer, output subsets, and weights are fixed before collecting the certification data.
We assume independent experimental runs with the same input state, a generic family of output subsets, and positive weights satisfying
$\sum_{\alpha=1}^q w_\alpha=1$.
The interferometer parameters are treated as known.
All runs are included, without conditioning on the number of detected particles.

For a single output subset and $r=2$, the observable in Eq.~\eqref{Zi} is
\begin{equation}
Z_i:=
\frac{\binom{N^{(i)}}{2}}{\binom{n}{2}}
=
\frac{N^{(i)}(N^{(i)}-1)}{n(n-1)},
\label{eq:pair_observable}
\end{equation}
where $N^{(i)}$ is the number of particles detected in the subset in run $i$.
Since $0\le Z_i\le1$, the unbiased estimator
\begin{equation}
\widehat P^{(2|n)}(R)
:=
\frac1R\sum_{i=1}^R Z_i
\end{equation}
satisfies Hoeffding's inequality~\cite{ProbIneq1963}:
\begin{equation}
\Pr{}_{\mathrm{data}}\left(
\left|\widehat P^{(2|n)}(R)-P^{(2|n)}\right|
\ge b
\right)
\le 2e^{-2Rb^2},
\qquad b>0.
\label{eq:hoeffding_pair}
\end{equation}
Here $\Pr_{\mathrm{data}}$ refers to the experimental outcomes for the fixed interferometer.

For a prescribed failure probability $0<\alpha<1$, define
\begin{equation}
b_R:=
\sqrt{\frac{\ln(2/\alpha)}{2R}}.
\label{eq:probability_precision}
\end{equation}
The estimation error is then at most $b_R$ with probability at least $1-\alpha$.

The same error bound applies to the weighted estimator
\begin{equation}
\widehat P_T^{(2|n)}(R)
:=
\frac1R\sum_{i=1}^R
\sum_{\alpha=1}^q w_\alpha
\frac{N_\alpha^{(i)}(N_\alpha^{(i)}-1)}
     {n(n-1)},
\end{equation}
where $N_\alpha^{(i)}$ is the number of particles detected in
$\Omega_\alpha$ in run $i$.
Indeed, each run contributes a convex combination of quantities in $[0,1]$, and hence a value in the same interval.
Independence across runs is sufficient; independence between counts in different subsets within a run is not required.
Thus,
\[
\Pr{}_{\mathrm{data}}\left(
\left|
\widehat P_T^{(2|n)}(R)-P_T^{(2|n)}
\right|\le b_R
\right)\ge1-\alpha.
\]

Estimate the probability defect in Eq.~\eqref{DeltP} by
\begin{equation}
\Delta_R:=
\begin{cases}
P_{T\max}^{(2|n)}-\widehat P_T^{(2|n)}(R),
&\text{bosons},\\[2pt]
\widehat P_T^{(2|n)}(R)-P_{T\min}^{(2|n)},
&\text{fermions}.
\end{cases}
\end{equation}
Since
\[
\left|\Delta_R-\Delta P_T^{(2|n)}\right|
=
\left|
\widehat P_T^{(2|n)}(R)-P_T^{(2|n)}
\right|,
\]
the quantity
\begin{equation}
\Delta_R^{\mathrm{up}}
:=
\max\{0,\Delta_R+b_R\}
\end{equation}
is an upper confidence bound on the true defect, with confidence at least $1-\alpha$.

Combining this bound with Eq.~\eqref{Dboudgap} gives
\begin{equation}
\mathcal D\ge
\mathcal D_{\mathrm{cert},R}
:=
\max\left\{
0,\,
1-\frac{n(n-1)}{2g(L_\omega)}
\Delta_R^{\mathrm{up}}
\right\}
\label{eq:finite_sample_certificate}
\end{equation}
with probability at least $1-\alpha$.
The clipping ensures that the reported bound lies in $[0,1]$ for every data set.

Before clipping, the estimator
$1-\frac{n(n-1)}{2g(L_\omega)}\Delta_R$
of the deterministic lower bound in Eq.~\eqref{Dboudgap} has statistical half-width
\begin{equation}
\epsilon_R
=
\frac{n(n-1)}{2g(L_\omega)}
\sqrt{\frac{\ln(2/\alpha)}{2R}}.
\end{equation}
Consequently,
\begin{equation}
R\ge
\frac{n^2(n-1)^2}
     {8g^2(L_\omega)\epsilon^2}
\ln\left(\frac2\alpha\right)
\label{eq:sample_complexity}
\end{equation}
suffices for a statistical half-width of at most $\epsilon>0$.
This controls the uncertainty in estimating the lower bound, rather than its difference from the actual value of $\mathcal D$.

For $0<\epsilon<1$, the certificate in
Eq.~\eqref{eq:finite_sample_certificate} reaches $1-\epsilon$ if and only if
\begin{equation}
\Delta_R+b_R
\le
\frac{2g(L_\omega)\epsilon}{n(n-1)}.
\label{certification_condition}
\end{equation}
For an indistinguishable input, the true defect vanishes.
On the concentration event
$|\widehat P_T^{(2|n)}(R)-P_T^{(2|n)}|\le b_R$,
we then have
$\Delta_R+b_R\le2b_R$.
Therefore, Eq.~(\ref{eq:sample_complexity}) 
 ensures that Eq.~\eqref{certification_condition} holds with probability at least $1-\alpha$, reproducing Eq.~\eqref{Runs}.

\subsection{Inverse-polynomial sensitivity for Haar-random interferometers}
\label{appC1}

We now consider a single output subset $\Omega_K$, with
$q=1$ and $w_1=1$, and define
\[
\omega_{\min}:=
\min_{k<l}\omega_{kl}
=
\min_{k<l}|H_{kl}|^2,
\]
where the minimum is taken over occupied input ports.

Let $U$ be Haar-distributed on $\mathrm U(M)$.
Fix $n$ occupied input ports, with $2\le n\le M$ and one particle per port, and a subset of $1\le K<M$ output ports, independently of $U$.
By Haar invariance, these inputs and outputs may be labeled
$1,\ldots,n$ and $1,\ldots,K$, respectively.
Throughout this subsection, $\Pr_U$ denotes probability over the Haar-distributed interferometer.
The matrix
$H=U\operatorname{diag}(I_K,0)U^\dag$
is a random rank-$K$ orthogonal projector.

Define the rising factorial by
$(x)_r:=x(x+1)\cdots(x+r-1)$ for $r\ge1$, with $(x)_0:=1$.
For every integer $r\ge1$, the character expansion of the unitary Weingarten function~\cite{collins2006} gives
\[
\sum_{\pi\in S_r}W_M(\pi)=\frac1{(M)_r}.
\]
Here the character sum defining $W_M$ is restricted to partitions of $r$ with at most $M$ parts, so the identity also applies when $r>M$.

To evaluate the moments of $H_{kk}$, take all input indices equal in Eq.~\eqref{AvH}.
For permutation operators acting on $(\mathbb C^K)^{\otimes r}$,
$\operatorname{Tr}\hat P_\tau=K^{c(\tau)}$,
where $c(\tau)$ is the number of cycles of $\tau$.
Since $(r!)^{-1}\sum_{\tau\in S_r}\hat P_\tau$ projects onto
$\operatorname{Sym}^r(\mathbb C^K)$,
\[
\begin{aligned}
\sum_{\tau\in S_r}K^{c(\tau)}
&=
\operatorname{Tr}\left(
\sum_{\tau\in S_r}\hat P_\tau
\right)\\
&=
r!\,\dim\operatorname{Sym}^r(\mathbb C^K)
=
r!\binom{K+r-1}{r}
=
(K)_r.
\end{aligned}
\]
It follows that
\begin{equation}
\int dU\,H_{kk}^r
=
\sum_{\tau\in S_r}K^{c(\tau)}
\sum_{\pi\in S_r}W_M(\pi)
=
\frac{(K)_r}{(M)_r}.
\label{eq:haar_diagonal_moments}
\end{equation}
These are the moments of the $\operatorname{Beta}(K,M-K)$ distribution.
A probability measure on $[0,1]$ is uniquely determined by its moments, so
\[
H_{kk}\sim\operatorname{Beta}(K,M-K).
\]

For $K=1$, write
$p_k:=|U_{k1}|^2$.
Then
\[
|H_{kl}|=\sqrt{p_kp_l},
\qquad
\sqrt{\omega_{\min}}
\ge\min_{1\le k\le n}p_k.
\]
Each $p_k$ has distribution $\operatorname{Beta}(1,M-1)$, and hence
\[
\Pr_U(p_k<t)
=
1-(1-t)^{M-1}
\le(M-1)t,
\qquad 0\le t\le1.
\]
The same argument applies to $K=M-1$ using the excluded column:
for $k\ne l$,
$H_{kl}=-U_{kM}U_{lM}^*$,
so we take $p_k:=|U_{kM}|^2$.

For $2\le K\le M-2$, set $\xi:=H_{kk}$.
The projector identity $H^2=H$ gives
\[
\sum_{l\ne k}|H_{kl}|^2=\xi(1-\xi).
\]
The distribution of $H$ is invariant under unitary conjugation.
Conjugations that fix the $k$th basis vector preserve $\xi$ and rotate the off-diagonal part of the $k$th row.
Consequently, conditional on $\xi$, the normalized vector
\[
\frac{(H_{kl})_{l\ne k}}{\sqrt{\xi(1-\xi)}}
\]
is uniformly distributed on the unit sphere in
$\mathbb C^{M-1}$.
Its squared coordinate moduli therefore have distribution
$\operatorname{Beta}(1,M-2)$, equivalently by applying
Eq.~\eqref{eq:haar_diagonal_moments} in dimension $M-1$ with rank one:
\[
\left.
\frac{|H_{kl}|^2}{\xi(1-\xi)}
\,\right|_{\xi}
\sim\operatorname{Beta}(1,M-2).
\]

For $Y\sim\operatorname{Beta}(1,M-2)$,
$\Pr(Y<s)\le(M-2)s$ for all $s\ge0$.
Conditioning on $\xi$ and then averaging gives
\begin{equation}
\begin{aligned}
\Pr_U(|H_{kl}|<t)
&\le
(M-2)t^2
\,\mathbb E_U\left[
\frac1{\xi(1-\xi)}
\right]\\
&=
\frac{(M-1)(M-2)^2}
     {(K-1)(M-K-1)}\,t^2.
\end{aligned}
\label{eq:haar_entry_tail}
\end{equation}
Here the Beta distribution of $\xi$ and
$[\xi(1-\xi)]^{-1}=\xi^{-1}+(1-\xi)^{-1}$ give
\[
\mathbb E_U\left[
\frac1{\xi(1-\xi)}
\right]
=
\frac{M-1}{K-1}
+
\frac{M-1}{M-K-1}.
\]

A union bound over occupied input ports in the boundary cases, and over pairs of occupied input ports in the interior case, yields
\begin{equation}
\Pr_U\left(\sqrt{\omega_{\min}}<t\right)
\le
\begin{cases}
n(M-1)t,
&K=1\ \text{or}\ K=M-1,\\[3pt]
\displaystyle
\binom n2
\frac{(M-1)(M-2)^2}
     {(K-1)(M-K-1)}\,t^2,
&2\le K\le M-2.
\end{cases}
\label{eq:haar_minimum_tail}
\end{equation}
No independence between matrix entries is required.

Set $t=\delta/[n(M-1)]$, where $0<\delta<1$.
For $K=1$ or $K=M-1$, the upper bound in
Eq.~\eqref{eq:haar_minimum_tail} equals $\delta$.
In the interior case, necessarily $M\ge4$, we use
\[
\frac{n-1}{n}\le\frac{M-1}{M},
\qquad
(K-1)(M-K-1)\ge M-3,
\]
to obtain
\[
\Pr_U\left(\sqrt{\omega_{\min}}<t\right)
\le
\frac{(M-2)^2}{2M(M-3)}\,\delta^2
\le
\frac{\delta^2}{2}
\le\delta.
\]
Thus, for every $1\le K<M$,
\begin{equation}
\Pr_U\left[
\sqrt{\omega_{\min}}
\ge\frac{\delta}{n(M-1)}
\right]\ge1-\delta.
\end{equation}

Choosing $\delta=(nM)^{-c}$ for fixed $c>0$ gives
\[
\omega_{\min}^{-1/2}
\le
n(M-1)(nM)^c
\le
(nM)^{c+1}
\]
with probability at least $1-(nM)^{-c}$.
On the same event, Eq.~\eqref{glowb} implies
\[
\frac{n}{g(L_\omega)}
\le
\omega_{\min}^{-1}
\le
(nM)^{2c+2}.
\]
This establishes the inverse-polynomial sensitivity and the polynomial sampling bounds stated in Section~\ref{Sec6B}. 
\section{Visibility and sampling cost: marginal versus full bunching}
\label{appD}

We compare the two-particle marginal bunching statistic ($r=2$) with the full $n$-particle generalized bunching statistic ($r=n$) of Ref.~\cite{GenBunch2016}, allowing each method to use a different number of selected output ports. We first consider lossless Haar-random unitary interferometers. Throughout this appendix, an overbar denotes the Haar average. The subscript  ``B'' refers to indistinguishable bosons, with $J(\sigma)\equiv1$, and ``d'' to distinguishable particles, with $J(\sigma)=\delta_{\sigma,I}$.

  We assume one particle per occupied input port, with the occupied inputs and the selected output subset fixed independently of $U$.

\subsection{A common visibility criterion}

Define the relative contrast by
\begin{equation}
\mathcal V_r(K):=
\frac{\overline P_{\mathrm B}^{(r|n)}(K)
      -\overline P_{\mathrm d}^{(r|n)}(K)}
     {\overline P_{\mathrm d}^{(r|n)}(K)}.
\label{appD:visibility}
\end{equation}
This quantity is the relative difference between the Haar-averaged probabilities. Normalizing the same difference by the indistinguishable-boson mean probability instead gives $\mathcal V_r/(1+\mathcal V_r)$.

\subsubsection{The two-particle marginal bunching statistic}

The two-particle averages are
\be
\overline P_{\mathrm B}^{(2|n)}(K)
=
\frac{K(K+1)}{M(M+1)},
\qquad
\overline P_{\mathrm d}^{(2|n)}(K)
=
\frac{K(KM-1)}{M(M^2-1)}.
\en{appD:pair_means}
Consequently,
\begin{equation}
\mathcal V_2(K)=\frac{M-K}{KM-1}.
\label{appD:pair_visibility}
\end{equation}
For any prescribed contrast $0<v\le1$, independent of $n$ and $M$,
\begin{equation}
\mathcal V_2(K)\ge v
\quad\Longleftrightarrow\quad
K\le\frac{M+v}{vM+1}.
\label{appD:pair_subset}
\end{equation}
Thus the prescribed contrast can be retained with $K=\mathcal O(1)$, for which
$\overline P_{\mathrm B}^{(2|n)}=\Theta(M^{-2})$.
In particular, $K=1$ gives $\mathcal V_2=1$ and
$\overline P_{\mathrm B}^{(2|n)}=2/[M(M+1)]$.
For a single selected output port, the contrast also equals one for every individual interferometer for which the distinguishable-particle probability is nonzero, since $|H_{kl}|^2=H_{kk}H_{ll}$.

\subsubsection{The full $n$-particle bunching statistic}

For full bunching, the corresponding expressions are
\be
\overline P_{\mathrm B}^{(n|n)}(K)
=
\frac{(K)_n}{(M)_n},
\qquad
\overline P_{\mathrm d}^{(n|n)}(K)
=
\left(\frac{K}{M}\right)^n
\left[1+\mathcal O\!\left(\frac{n^2}{KM}\right)\right],
\en{appD:full_means}
where $(x)_n=x(x+1)\cdots(x+n-1)$.
The first identity is exact; the second is the asymptotic estimate of Ref.~\cite{GenBunch2016}.
For $K,M\gg n$ and $n^2/(KM)\ll1$, expanding the logarithm of the ratio of these averages gives
\be
\ln[1+\mathcal V_n(K)]
=
\frac{n(n-1)}{2}\left(\frac{1}{K}-\frac{1}{M}\right)
+
\mathcal O\!\left(
\frac{n^3}{K^2}+\frac{n^3}{M^2}+\frac{n^2}{KM}
\right).
\en{appD:full_visibility}

Consider the asymptotic boson-sampling regime
$M\ge c_0n^2$, where $c_0>0$ is fixed and $M$ grows at most polynomially with $n$.
Let $\ell_v:=\ln(1+v)>0$, and let $K_v$ denote the largest subset size compatible with $\mathcal V_n(K_v)\ge v$.
Solving the contrast condition to leading order gives
\begin{equation}
K_v
=
\frac{n(n-1)}
     {2\ell_v+\frac{n(n-1)}{M}}
\left[1+\mathcal O(n^{-1})\right].
\label{appD:full_subset}
\end{equation}
Thus $K_v=\Theta(n^2)$, which makes the expansion in Eq.~\eqref{appD:full_visibility} self-consistent: its remainder is $\mathcal O(n^{-1})$ in this regime.
Substituting into the exact bosonic mean yields
\begin{equation}
\ln\overline P_{\mathrm B}^{(n|n)}(K_v)
=
-n\ln Q_v+\mathcal O(1),
\qquad
Q_v:=1+\frac{2M\ell_v}{n(n-1)}.
\label{appD:exponential_probability}
\end{equation}
Since $Q_v\ge1+2c_0\ell_v>1$, the mean probability decays at least exponentially with $n$.
The exact bosonic mean increases with $K$, so choosing fewer output ports only decreases this mean.
Consequently, optimizing the subset size subject to a fixed nonzero ensemble contrast still leaves an exponentially small full-bunching mean probability. The two-particle marginal statistic retains the prescribed contrast with an inverse-polynomial mean probability.

\subsection{Sampling consequences at fixed contrast}

We now relate this comparison to sampling cost using the bounded marginal observable defined in Eq.~\eqref{eq:pair_observable}.
\subsubsection{The two-particle marginal bunching}

On a fixed
interferometer, a run with $N_K=|\bm{m}|_K$ particles detected in the selected
subset $\Omega_K$ supplies the bounded random variable
\begin{equation}
 Z=\frac{\binom{N_K}{2}}{\binom n2},
 \qquad 0\le Z\le1,
 \qquad \mathbb E Z=P^{(2|n)}.
 \label{appD:pair_statistic}
\end{equation}
The difference between its means for the two limiting inputs is
\begin{equation}
 \gamma(U):=P_{\mathrm B}^{(2|n)}-P_{\mathrm d}^{(2|n)}
 =\binom n2^{-1}\sum_{k<l}|H_{kl}|^2
 \ge\omega_{\min}.
 \label{appD:pair_gap}
\end{equation}
Assuming the interferometer is calibrated, a test with threshold
halfway between these two means has error probability at most
$\alpha$ under either hypothesis if
\begin{equation}
 R\ge \frac{2}{\gamma^2(U)}\ln\frac{2}{\alpha}.
 \label{appD:pair_runs}
\end{equation}
This sufficient bound follows directly from Hoeffding's
inequality. By Appendix~\ref{appC}, with probability at least
$1-\delta$ over the interferometer,
\begin{equation}
 \omega_{\min}^\frac12\ge\frac{\delta}{n(M-1)}.
 \label{appD:haar_sensitivity}
\end{equation}
It therefore suffices to choose
\begin{equation}
 R\ge
 2\left[\frac{n(M-1)}{\delta}\right]^4
 \ln\frac{2}{\alpha}.
 \label{appD:polynomial_runs}
\end{equation}
  
\subsubsection{The full $n$-particle bunching}
For the full-bunching statistic, each run instead supplies
$Z=\delta_{N_K,n}$. Write
$p_{\mathrm B}(U)=P_{\mathrm B}^{(n|n)}(U)$ and
$p_{\mathrm d}(U)=P_{\mathrm d}^{(n|n)}(U)$.
For these two limiting inputs,
$0\le p_{\mathrm d}(U)\le p_{\mathrm B}(U)$.
The total variation distance between their distributions of
$R$ independent full-bunching indicators is at most
$R|p_{\mathrm B}(U)-p_{\mathrm d}(U)|\le R p_{\mathrm B}(U)$.
Consequently, any test using these indicators, with error at most
$\alpha<1/2$ under each hypothesis, necessarily satisfies
\begin{equation}
 R\ge\frac{1-2\alpha}{p_{\mathrm B}(U)}.
 \label{appD:full_runs_lower}
\end{equation}
Moreover, Markov's inequality gives
\begin{equation}
 \Pr_U\!\left[
 p_{\mathrm B}(U)\le
 \frac{\overline P_{\mathrm B}^{(n|n)}}{\delta}
 \right]\ge1-\delta.
 \label{appD:rare_event_bound}
\end{equation}
Combining Eqs.~\eqref{appD:exponential_probability},
\eqref{appD:full_runs_lower}, and
\eqref{appD:rare_event_bound} shows that, with probability at
least $1-\delta$, the full-bunching indicator requires
\begin{equation}
 R\ge
 \frac{\delta(1-2\alpha)}
      {\overline P_{\mathrm B}^{(n|n)}}
 =\delta(1-2\alpha)\exp\!\left[n\ln Q_v+\mathcal O(1)\right]
 \label{appD:exponential_runs}
\end{equation}
runs at the fixed ensemble contrast. This remains exponential when
$\delta$ decreases only inverse-polynomially.

The separation above relies on keeping the contrast bounded away from zero as $n$ increases. In the regime considered above, taking $K=M-L$ with $L\sim cM/n$ and fixed $c>0$ gives, as $n\to\infty$,
\begin{equation}
\overline P_{\mathrm B}^{(n|n)}\longrightarrow e^{-c},
\qquad
\mathcal V_n\sim\frac{c(n-1)}{2M}.
\label{appD:vanishing_visibility}
\end{equation}
The mean event probability then tends to a positive constant, while the contrast vanishes.
The sampling advantage established here therefore applies to fixed nonvanishing contrast; it does not imply a universal ordering of run counts when the contrast is unconstrained.

\subsection{Comparison under  uniform losses}
Finally, let $0<\eta\le1$ denote the uniform single-particle survival probability. The unconditioned marginal probabilities obey
\begin{equation}
P_X^{(r|n)}(\eta)=\eta^rP_X^{(r|n)}(1),
\qquad
X\in\{\mathrm B,\mathrm d\},\quad r=2,n.
\label{appD:uniform_loss}
\end{equation}
Uniform losses leave the relative contrasts unchanged and multiply the two-particle and full-bunching probabilities by $\eta^2$ and $\eta^n$, respectively. For fixed $0<\eta<1$, this introduces an additional exponential suppression of the full-bunching probability.
The difference between the two pair-statistic means is also multiplied by $\eta^2$, so the sufficient run bound in Eq.~\eqref{appD:pair_runs} increases by a factor of $\eta^{-4}$. Polynomial sampling cost is therefore retained whenever $\eta^{-1}$ grows at most polynomially with $n$ and $M$.
These statements apply to statistics formed from all input trials, without conditioning on the survival of all particles.
 
 \bibliography{references}

 \end{document}